# Strain Wave Pathway to Semiconductor-to-Metal Transition revealed by time resolved X-ray powder diffraction

C. Mariette[1,*], M. Lorenc[1,*], H. Cailleau[1], E. Collet[1], L. Guérin[1], A. Volte[1], E. Trzop[1], R. Bertoni[1], X. Dong[1], B. Lépine[1], O Hernandez[2], E. Janod[3], L. Cario[3], V. Ta Phuoc[4], S. Ohkoshi[5], H. Tokoro[5,6], L. Patthey[7], A. Babic[7], I. Usov[7], D. Ozerov[7], L. Sala[7], S. Ebner[7], P. Böhler[7], A Keller[7], A. Oggenfuss[7], T. Zmofing[7], S. Redford[7], S. Vetter[7], R. Follath[7], P. Juranic[7], A. Schreiber[7], P. Beaud[7], V. Esposito[7,a)], Y Deng[7], G. Ingold[7], M. Chergui[8], G. F. Mancini[7,8], R. Mankowsky[7], C. Svetina[7], S. Zerdane[7], A. Mozzanica[7], M. Wulff[9], M. Levantino[9], H. Lemke[7], M. Cammarata[1,b)*]

1 Univ Rennes, CNRS, IPR (Institut de Physique de Rennes) - UMR 6251, F-35000 Rennes, France

2 Univ Rennes, CNRS, ISCR (Institut des Sciences Chimiques de Rennes) - UMR 6226, F-35000 Rennes, France

3 Institut des Matériaux Jean Rouxel (IMN), Université de Nantes, CNRS, 2 rue de la Houssinière, F-44322 Nantes, France

4 GREMAN – UMR 7347 CNRS, Université de Tours, Parc de Grandmont, F-37000 Tours, France.

5 Department of Chemistry, School of Science, The University of Tokyo, 7-3-1 Hongo, Bunkyo-ku, Tokyo, Japan

6 Department of Materials Science, Faculty of Pure and Applied Sciences, University of Tsukuba, 1-1-1 Tennodai, Tsukuba, Ibaraki, Japan

7 SwissFEL, Paul Scherrer Institut, 5232 Villigen PSI, Switzerland

8 Laboratory of Ultrafast Spectroscopy, Lausanne Center for Ultrafast Science (LACUS), École Polytechnique Fédérale de Lausanne,14 CH-1015 Lausanne, Switzerland

9 European Synchrotron Radiation Facility, 71, avenue desMartyrs, F-38000 Grenoble, France.

*to whom correspondence should be addressed: celine.mariette@univ-rennes1.fr, maciej.lorenc@univ-rennes1.fr, marco.cammarata@univ-rennes1.fr

a) current address: Stanford Institute for Materials and Energy Science , Stanford University and SLAC National Accelerator Laboratory , Menlo Park, California 94025, USA

b) current address: European Synchrotron Radiation Facility, 71, avenue desMartyrs, F-38000 Grenoble, France.

## Abstract

Thanks to the remarkable developments of ultrafast science, one of today's challenges is to modify material state by controlling with a light pulse the coherent motions that connect two different phases. Here we show how strain waves, launched by electronic and structural precursor phenomena, determine a macroscopic transformation pathway for the semiconducting-to-metal transition with large volume change in bistable $Ti_3O_5$ nanocrystals. Femtosecond powder X-ray diffraction allowed us to quantify the structural deformations associated with the photoinduced phase transition on relevant time scales. We monitored the early intra-cell distortions around absorbing metal dimers, but also long range crystalline deformations dynamically governed by acoustic waves launched at the laser-exposed $Ti_3O_5$ surface. We rationalize these observations with a simplified elastic model, demonstrating that a macroscopic transformation occurs concomitantly with the propagating acoustic wavefront on the picosecond timescale, several decades earlier than the subsequent thermal processes governed by heat diffusion.

Unprecedented opportunities have emerged for transforming a material with a femtosecond (fs) laser pulse [**Fleming2008**, **Nasu2004, Yonemitsu2008, Zhang2014, Basov2017**]. New breakthroughs are anticipated, such as attaining a macroscopic transformation by driving the material on a deterministic pathway from one phase to another, ideally in a fast and highly efficient fashion. During these transformations different degrees of freedom couple sequentially and give rise to multiscale dynamics in space and time. Some of these degrees of freedom result in structural reorganisations that are crucial for stabilizing a photoexcited electronic state. Many studies have focused so far on optical coherent phonon oscillations around new atomic positions [***Johnson2017***, **Beaud2014**]. However, this process drives intracell atomic displacements that preserve the crystal shape and volume. Stabilization of the new macroscopic structural order requires the establishment of long range crystal deformations involving propagating acoustic waves. In a recent study on spin-crossover (SCO) molecular crystals we showed that cooperative interactions of elastic origin help attain a macroscopically robust switched state [**Bertoni2016**]. In another study of photo-induced insulator to metal transition, it was proposed that the local expansion of the lattice drives the propagation of the metallic region [**Okimoto2009**]. Suspecting key role of the lattice strain in photo-induced phase transition [**Wang2019**], we investigate the semiconducting-to-metallic phase transition in Ti$_3$O$_5$ nanocrystals which exhibits even greater volume change on transition than the above examples, and high resilience to laser flux. Importantly, in the aforementioned studies [**Bertoni2016, Okimoto2009**], transient optical spectroscopy was used to follow the photo-induced changes. Unfortunately optical observables are hard to relate to actual structural changes, in particular lattice deformations. In this contribution we use X-ray pulses from an X-ray Free Electron Laser (XFEL) and Synchrotrons to perform time-resolved powder diffraction, allowing a quantitative refinement of the evolving structures on fs to microsecond time scales. The ultrafast strain dynamics have been further rationalized by adopting a well-known phenomenological elastic model [**Thomsen1986a, Schick2014**] and accounting for phase transformation. The highly sensitive X-ray probe revealed significant and physically sound structural changes acting as precursor stress for the macroscopic phase switching, found to move with a strain wave at the speed of sound, and hereafter referred to as acoustic front.

The trititanium pentoxide (Ti$_3$O$_5$, Fig. 1) undergoes a thermal phase transition between a semiconducting (so-called $\beta$) and a metallic (so-called $\lambda$) phase around T$_{SM}$ = 460 K (upon heating) [**Onoda1998, Okhoshi2010**]. This semiconducting to Metal (S-M) phase transition is isostructural [**Zacharias2015**] (i.e. same monoclinic space group *C2/m* for both phases, Fig. 1a), strongly first order and characterized by significant changes in volume (+ 6.4 %) and latent heat (120 meV/unit) [**Tokoro2015**]. This increase in volume is mainly caused by the expansion of the lattice along the *c* crystalline axis. As shown in Fig. 1 the S-M phase transition is related to large intra-cell structural reorganization and the dissociation of the *Ti$_3$-Ti$_3$* dimers sharing electrons lying on a band just below the Fermi level in the $\beta$-phase (Fig. 1a and b). From the electronic standpoint, the S-M phase transition is therefore characterized by a vanishing electronic gap related to the dissociation of such *Ti$_3$-Ti$_3$* dimers [**Okhoshi2010, Kobayashi2017**]. A striking feature of this transition is that reducing the crystallite size has a spectacular influence on the metastability region of the $\lambda$-phase. For nano-sized crystallites, the $\lambda$-phase is stable down to room temperature (RT), effectively making the system bistable in a broad temperature range [**Ohkoshi2010**]. In the present study, we focus on a Ti$_3$O$_5$ pellet of nanocrystals containing 72.5 % $\beta$ phase and 27.5 % residual metastable $\lambda$ phase at room temperature (weight percentages, see Methods). The Ti$_3$O$_5$ nanocrystals have a typical size of about 100 nm and a morphology similar to the "*neat-flakes form*" described in [**Ohkoshi2010**]. X-Ray diffraction patterns were recorded for this pellet from 300K to 700 K. Figs. 1 g and h show the evolution of the *c*-axis parameter and monoclinic angle $\phi$ of the different phases. As expected, the $\lambda$ and $\beta$ phases coexist from room temperature to T$_{SM}$ . At higher temperature (T$_c$ = 500 K) a second order phase transition occurs towards a high symmetry metallic phase (so-called $\alpha$, of orthorhombic space group *Cmcm*) [**Onoda1998**]. This transition is of ferroelastic nature, with the monoclinic angle $\phi$ locking to 90° at T$_c$. In contrast to the S-M transition, it is continuous and does not exhibit any volume discontinuity (see orange curve in Fig. 1h). The metastable $\lambda$-phase can be switched to

$\beta$ - phase by applying an external pressure ($P_{MS}$= 0.5 GPa) whose value may strongly depend on the morphology [**Tokoro2015**, **Ohkoshi2019**]. Also a photo-reversible persistent phase transition between $\beta$ and $\lambda$ phases was reported under intense ns irradiation [**Ohkoshi2010**, **Ashara2014**, **Ould2014**]. Such a photoresponsive phase change material with robust RT bistability is not only of fundamental interest but also holds promise for different technological applications, such as efficient optical or heat storage [**Tokoro2015**, **Okoshi2019**]. The photoinduced $\beta$-to-$\lambda$ transition was also investigated in the transient regime, i.e. below the excitation threshold for permanent switching [**Asahara2014**, **Tasca2017**]. Full scale dynamics, from fs to µs, was probed by ultrafast diffuse reflection spectroscopy on a nanogranular pellet of $Ti_3O_5$; those measurements have been interpreted in terms of nucleation and growth process of $\lambda$ regions [**Asahara2014**]. Unfortunately the previous studies have lacked either the time resolution [**Tasca2017**] or the structural sensitivity [**Asahara2014**] to confirm such a hypothesis and to investigate the role of strain wave propagation and volume change in stabilizing the photoinduced phase.

Here we use ultrafast X-ray diffraction (XRD) at SwissFEL beamline Bernina [**Ingold2019**] to investigate the mechanism of the photoinduced $\beta$-to-$\lambda$ transition on a $Ti_3O_5$ nanocrystals pellet. The experimental geometry is sketched in Fig. 2a. The X-ray photon energy was set to 6.6 keV and the grazing angle to 0.5°. Considering both X-ray absorption and pellet roughness, the estimated effective probed depth for this geometry was $z_p$ = 400 nm (Figs. S2 and S3). The 1.55 eV, 500 fs long pump laser beam impinged the sample at 10° incident angle. At this energy, the pump penetration depth calculated from the refractive index reported in [**Hakoe2017**] was $\xi_L$ = 65 nm. The measured time resolution was about 600 fs FWHM [**Ingold2019**]. These studies are complemented by experiments at the ID09 beamline at ESRF, probing longer time scale with time resolution around 100 ps (Fig. S4). Unless stated otherwise, we will refer to the SwissFEL experiment. Fig. 2b compares powder diffraction differential patterns taken at various time delays. The low noise, almost featureless transient at -4.5 ps, highlights the high quality of the differential data and data reduction procedures. The differential patterns in Fig. 2b emphasizes that major changes occur already at very early delays indicating rapid structural deformations within 500 fs. At picosecond time scale the difference signal increases in amplitude; the signal around some Bragg peaks changes towards a characteristic "bi-polar", i.e. positive change towards low angles and negative change towards higher angles indicating an increase unit-cell parameters. Broadening of some Bragg peaks is also observed ("butterfly shape"), indicating inhomogeneous lattice distortions; while significant changes in the absolute diffracted intensity suggest inter-atomic re-organisations. The high amplitude of the differential patterns (from 10 % to 30 % depending on the q region) indicates large photo-induced deformations/transformations taking place in the sample. Despite the complexity of the structural changes, the high-fidelity data allow a complete Rietveld analysis of the patterns for each time delay. Figs. 2c-d show the results of the Rietveld refinement of the diffraction patterns for a reference pattern (pump laser off) and for a pump / probe delay of 7.5 ps (see also Fig. S5). In both cases the Rietveld analysis with unconstrained cell and atomic position parameters for the $\lambda$- and $\beta$-phases yielded similar reliability factors ($R_p$= 3% and $R_{wp}$= 7-8%) and difference profile curves compared to the refinement of the pattern recorded under equilibrium condition. This attests of the quality of the Rietveld analysis at positive time delays and enables the discussion of the time evolution of the structure at the sub-ps time-scale, which we discuss hereafter. All details about the sample, experiment, data reduction and analysis are given in the Methods section).

We first describe the ultra-fast structural changes occurring within the first 500 fs where significant intra-cell distortions are already observed. The evolution of selected angles and distances are displayed in Fig. 3 for delays up to 7 ps, typical changes being in the order of $10^{-2} \pm 2 \times 10^{-3}$ Å and 0.5-1 ± 0.02 °. Comparison with calculated electronic structure (Fig. 1a to f) helps to rationalize the observed photoinduced changes. The calculated optical conductivity is particularly insightful for describing the possible electronic transitions induced by the 1.55 eV pump photons (Fig. 1e and f). For the majority $\beta$-phase, the two bands lying below and closest to the Fermi level centered at 1.1 eV and 0.3 eV have a strong $Ti_2$ and $Ti_3$ character, respectively.

Although both $Ti_2$-$Ti_2$ and $Ti_3$-$Ti_3$ dimers are excitable above the gap by 1.55 eV photons, the calculated optical conductivity is mainly contributed in this range by excitation from $Ti_3$-$Ti_3$ dimers to the conduction band. Electron depletion of the bonding $Ti_3$-$Ti_3$ states thus leads to the observed fast increase of the $Ti_3$-$Ti_3$ distance (Fig. 3b). The $Ti_3$-$Ti_3$ dimer also rotates towards the isolated $Ti_1$ atom and away from the second ($Ti_2$-$Ti_2$) dimer (Fig. 3a). These motions of small amplitude are precursor structural signatures of the isostructural phase transition from the $\beta$ - towards the $\lambda$ - phase. However, the not-complete rotation (compared to thermal equilibrium) results in a significant decrease of the $Ti_2$-$Ti_2$ distance. On the same time scale, the bond length of the second dimer $Ti_2$-$Ti_2$ remains unchanged (Fig. 3b). For the minority $\lambda$-crystallite, the photoexcitation promotes the $Ti_2$ and $Ti_3$ electrons above the Fermi level. The contributions to the calculated optical conductivity from both $Ti_2$ and $Ti_3$ electrons around 1.55 eV are comparable (see Fig. 1h), albeit they lead to different structural effects. We observe a significant increase of $Ti_2$-$Ti_2$ distance on a sub-ps time scale (Fig. 3e): depletion of the $Ti_2$-$Ti_2$ bonding orbital weakens the $Ti_2$-$Ti_2$ dimer. The $Ti_2$-$Ti_2$ distance thus becomes closer to that of non-dimerized $Ti_3$-$Ti_3$, initially slightly longer. In the high symmetry $\alpha$ phase, $Ti_2$ and $Ti_3$ are equivalent through a mirror plane (Fig. 3h). The relative evolution of $Ti_2$-$Ti_2$ and $Ti_3$-$Ti_3$ distances (Fig. 3e), as well as $Ti_2$-$Ti_2$-$Ti_3$ and $Ti_2$-$Ti_3$-$Ti_3$ angles (Fig. 3d), directly probe the ultrafast evolution of the degree of symmetry breaking in the low symmetry monoclinic metallic phase. These precursor structural changes occur before long-range cell deformations are observed. In this early temporal window, the volume of the $\beta$ - phase is almost constant, while the volume of the $\lambda$ - phase has a linear increase with no discontinuity around zero time. This notwithstanding, the unit cell undergoes ultrafast shear. The decrease of the monoclinic angle ($\phi$) plotted in Fig. 3f indicates partial symmetry change within 4 ps, with a clear and prompt response. It is noteworthy that locking is incomplete at this stage ($\phi \sim 90.80°$). A high resolution diffuse scattering experiment (Fig. S6) revealed the existence of important stacking faults along the propagation of this ferroelastic distortion (**c** direction). This suggests that the macroscopic evolution of the monoclinic shear occurs independently on coherent domains with a size much smaller than the nanocrystal (few unit cells), as schematically depicted in Fig. S6.

The evolution of the $\lambda$ - phase unit cell volume takes place on a longer time scale ($\Delta V_\lambda$, Fig. 4a). The increase is almost linear up to ≈ 16 ps, the maximum of Bragg peak broadening resulting from the strain distribution, roughly quantified here by a microstrain parameter ($\varepsilon$) conventionally used in powder diffraction analysis. The evolution of the $\lambda$ - macroscopic phase fraction ($\Delta X_\lambda$, Fig. 4c), signals that part of the $\beta$ - crystallites undergoes the phase transition. At around 20 ps, $\lambda$ - phase ratio has increased from 27.5 % to 33.0 ± 0.8 % on average within the X-ray probed depth (400 nm). Importantly its evolution is quasi linear up to t ≈ 16 ps, following essentially the dynamics of the $\lambda$ - phase unit cell volume. The behaviour of the $\beta$-phase unit cell volume is less straightforward. Normalized evolution ($\Delta V_\beta$, Fig. 4c) reveals a small decrease to a minimum around t ≈ 16 ps. Subsequently, the volume starts to increase to level the equilibrium value and well above. The latter changes remain small, the volume of this $\beta$ - phase increases from 349.3 to 350.7 Å$^3$ (+0.4%), while it increases from 371.7 to 381.3 ± 0.1 Å$^3$ (+2.5%) for the metallic phase.

The connection between ultrafast photo-excitation and strain dynamics leading to the overall volume change has been described by Thomsen [**Thomsen1986a**] and later extensively used to describe strain propagation [**Schick2014**, **Ruello2015**]. We performed numerical calculations for $\lambda$ - and $\beta$ - crystallites separately, in order to rationalize our observations (see SI, Fig. S7). The initial stress profile in the crystallites is assumed to follow the exponential profile of the laser penetration. To establish the mechanical equilibrium between these crystallites and their environment, acoustic wave packets propagate from the surface to interior, leaving behind a static deformation. In the present case however, stress arises primarily from the precursor photoinduced electronic changes and structural distortions described in the previous paragraphs, unlike the original model which considered laser induced thermal stress. The occurrence of a phase transition raises the complexity of description. The $\lambda$-crystallites, despite undergoing a phase transition to the $\alpha$ - phase, have no volume anomaly detected at T$_c$, so that the strain dynamics triggered at t$_0$ can be described as a continuous thermal expansion. The mean volume change is derived by integrating the strain. The strain gradient, arising from the initial stress distribution, can be quantified by the standard deviation of the calculated strain (Fig. 4b). It can then be compared with the observed Bragg peak broadening, quantified in the Rietveld refinement by the microstrain parameter epsilon (Fig. 4a). The observed clear anomaly at 16 ps in the microstrain dynamics sets up the acoustic time scale $\tau_{ac}$. A good agreement for the macroscopic parameters considered here is found with the model for wavefront propagating on L = 100 nm at sound velocity of $6.5 \times 10^3$ m.s$^{-1}$, independently determined with picosecond interferometry [**Thomsen1986b**], sensitive to propagating acoustic wavepackets [**Asahara2014**] (Fig. S8). In the same framework, the 4 ps shear dynamics described above, assuming a transverse acoustic velocity around $3 \times 10^3$ m.s$^{-1}$ [**Tokoro2015**], would correspond to a propagation of about 10 nm. Ferroelastic domains of similar size were recently reported for an archetypal oxide [**Singer2018**]. For the $\beta$-phase, we accounted for a partial transformation of the crystallites, and the associated unit cell volume changes. The initially excited $\beta$-crystallites are assumed to be promptly transformed to $\lambda$-phase by the moving acoustic front, as depicted in Fig. 4 e, and no additional strain upon feedback from the phase transition was factored in the calculations. The resulting phase propagation and the associated long range deformation are schematically depicted in Fig. 4 e. The phase front is assumed to stop at L = 100 nm, matching the average crystallite size as observed by X-ray diffraction. The acoustic front takes about 16 ps to travel this distance (see black line in Fig. 4 a - d). This is exactly when the $\beta$-phase volume shows a well defined minimum, a direct consequence of the compression exerted by the relatively high volume of the layer transformed to the $\lambda$-phase (Fig. S7). The latter also influences the calculated mean strain for $\lambda$ (Fig. 4 b). The calculated $\lambda$-phase fraction is plotted in Fig. 4 d and it reproduces very accurately the behavior in Fig. 4 c . As a consequence, the increase of the latter by 5% on average within the estimated X-ray probed depth, amounts to approximately 26% within the first 100 nm below the surface, where the phase front stops (see SI, Fig. S9).

On longer time-scales (t> $\tau_{ac}$), there subsist stationary temperature and strain distributions [**Thomsen1986a, Schick2014**]. The recovery to a homogeneous state requires slower heat diffusion inside the sample. This process was investigated by XRD with 100 ps time resolution [**Cammarata2009**] (Fig. 5a). The slow process can be clearly distinguished because it develops until 100 ns. Temperature homogenization leads to an increase of the average sample temperature, and consequently the transformation of the probed $\beta$ - crystallites reaches 30 % in the probed volume (Fig. 5b). The 100 ns value is consistent with expected time for heat diffusion over D ≈ 200 nm (in a bulk material with thermal diffusivity of 230 nm$^2$.ns$^{-1}$) [**Tokoro2015**]. The complete recovery of thermal equilibrium with the environment is observed on a 10 µs time scale, in agreement with ref. [**Tasca2017**].

In summary, this work offers a new perspective for the ultrafast control of materials with evidence of a strain wave pathway for driving a phase transition in nanocrystallites excited by a light pulse. Key role of the long-range lattice deformation dynamics in the transformation process is highlighted, and so are the benefits of direct structural probes for visualising thereof. In particular, we show that state of the art XFEL sources permit measurements of the real-time interatomic motions and lattice distortions, even in a bi-phasic polycrystalline powder. Our experiment, corroborated with a phenomenological elastic model, unambiguously reveals a phase front starting from the sample surface and moving with sonic speed into the bulk. Notably, the strain waves are launched directly in the material by internal precursor stresses that store mechanical energy. This coherent process restoring mechanical equilibrium, takes place on ultrafast timescale and clearly precedes the much slower thermal homogenization. It is a step forward for self-contained transformation, compared to previously reported phase front propagation which required a hetero-structure for launching the strain waves at the inferface of a strongly absorbing transducer film into the phase change film [**Först2015, Först2017**]. All in all, the mechanism revealed herein is fundamentally different from the nucleation and growth, rooted in phase transitions at thermal equilibrium and so far proposed in a similar context [*Asahara2014*, *Singer2018*]. We believe that the strain wave pathway is likely to be valid in a variety of volume changing materials.

## Acknowledgements

The authors want to thank Gemma Newby and Martin Pedersen from ESRF for support during earlier phases of this project, Alexei BOSSAK (ID28, ESRF) for carrying out diffuse scattering measurements discussed in SI, Guénolé Huitric (IPR) for his contribution to the optical spectroscopy measurements performed at IPR and discussed in SI and Roman Bertoni (IPR) for fruitful discussions and suggestions. MC wants to thank PSI for accepting his proposal to study $Ti_3O_5$ nanoparticles as the first experiment for the entire facility and for providing excellent support. MLo gratefully acknowledges Agence Nationale de la Recherche for financial support under grant ANR-16-CE30-0018 ("Elastica"). BL would like to thank Bruno Bêche (IPR) for help with AFM measurement. MC and MLe acknowledge the support of European Union Horizon 2020 under the Marie Sklodowska-Curie Project "X-Probe" Grant No. 637295. SO and HT acknowledge support from JSPS Grant-in-Aid for specially promoted Research Grant Number 15H05697, and JSPS KAKENHI Grant Number 16H06521 Coordination Asymmetry. MCh and GFM acknowledge funding from the European Research Council (ERC) H2020 DYNAMOX (grant No 695197 and from the swiss NSF via the NCCR:MUST). Parts of this research were carried out in the frame of the IM-LED LIA (CNRS).

## Author contributions

CM, MLo, HC and MC coordinated the project. HC and MLo proposed the study of the photoinduced phase transition of $Ti_3O_5$. MC coordinated the X-ray studies. CM developed the thermo-elastic model with the help of MLo. Nanocrystals of $Ti_3O_5$ have been characterized and synthesized by SO and HT. CM and LC performed Rietveld refinement with help from OH. OH performed the static powder XRD measurements. MC and MLe developed the setup at ESRF with help of MW, SZ, ET, EC, XD, CM, MLo. MC and HL developed the setup at SwissFEL with help from LG, AV, CM, MLe, LP, PB, VE, YF, GI, GM, RM, CS, SZ. MC developed the data reduction procedure. AM developed the Jungfrau detector used for the experiment. VT performed DFT calculations with help from EJ. CM, MLo, HC and MC wrote the manuscript with significant contributions from EC, EJ, LC, MLe, PB, critical reading from MCh and help from all authors.

# Methods

## Static powder laboratory X-ray diffraction - temperature measurements

The static powder XRD as a function of temperature was measured in Debye-Scherrer geometry on a Bruker AXS D8 Advance (Mo-K$_\alpha$ radiation selected with a focusing Göbel mirror) equipped with a MRI high temperature capillary furnace and a high-energy LynxEye detector. The flake form $Ti_3O_5$ powder sample [**Ohkoshi2010**] was sealed in a quartz capillary of 0.3 mm in diameter.

## DFT/optical properties calculations

All calculations have been carried out by using the Quantum ESPRESSO package within the framework of DFT+U, with the Perdew-Burke-Ernzerhoff (PBE) generalized gradient approximation (GGA) to describe the exchange-correlation functional. Both Projector Augmented Wave basis (PAW) and norm-conserving pseudo-potentials. The Monkhorst-Pack grid of 8x8x4 in the reciprocal space was used for the Brillouin zone sampling for both $\lambda$ - phase and $\beta$ - phase. The total energy of the system converged to less than $1.0 \times 10^{-6}$ Ry. Electronic wave functions were represented in a plane wave basis up to an energy cut-off of at 90 Ry. Crystallographic structures were taken from [**Okhoshi2010**]. Ferromagnetic and antiferromagnetic (AFM) orders were considered. AFM order was found to be the ground-state for both $\lambda$ - and $\beta$ -phases. The optical properties were computed using epsilon.x post-processing tool of the Quantum Espresso package, at the independent-particle approximation level. Both intraband and interband contributions were considered. Calculated optical density shown in Fig. 1 is the average of the 3 diagonal elements (Fig. S1).

## Samples preparation

Flakes-form $\lambda$ -phase samples were obtained following the synthesis method described in [**Ohkoshi2010**]. Pellets were made from flakes powder using a uniaxial press at 3 GPa. The resulting pellets have a density of 3.2 g.cm$^{-3}$ (a ratio of 0.8 compared to single crystal, measured by X-ray absorption). They contain a mixture of pressure-induced $\beta$ -phase crystallites and $\lambda$ -phase crystallites due to residual stress. The absolute $\lambda$ -phase fraction is determined by Rietveld refinement around 25-30% (depending on the pellet), in good agreement with the ratio usually observed with this method [**Ohkoshi2010**]. Whether $\beta$ - and $\lambda$ - could coexist within a single grain remains an open question and cannot be answered with our XRD measurements.

## Experimental details - time resolved experiment

Experiments were performed at ESRF Synchrotron (Grenoble, France) using the ID09 time resolved beamtime (**exp1**) and at SwissFEL (PSI, Villigen, Switzerland) at the Bernina beamline. The latter was part of the very first commissioning experiment of the entire facility (**exp2**).

### EXP1: ESRF experiment detail

The ID09 setup has been discussed in detail previously [**Cammarata09**]. Briefly, fast rotating choppers were used to isolate X-ray single pulses (each 100 ps long) at 1 kHz repetition rate. A fast shutter was used to lower the frequency to 10 Hz. The X-rays were partially monochromatized by using a Ru/B$_4$C multilayer monochromator resulting in 1.5% bandwidth centered at 11.5 keV and focused by a toroidal mirror to a size of 0.1 x 0.06 mm². In order to reduce the X-ray beam footprint on the sample, the last slits (~0.6 m from the sample) were closed vertically to a 0.03 mm gap. A synchronized laser (800 nm) at 10 Hz was used to excite the sample using a perpendicular geometry configuration, with the laser hitting the sample from the top. Laser Beam size was 10x0.23 mm² or 4.6x0.21 mm² depending on power densities (respectively below and above 1 mJ.mm$^{-2}$). Diffracted X-rays were integrated on a Rayonix

MX170-HS CCD detector and azimuthally integrated using pyFAI [*Ashiotis2015*].

**EXP2: SwissFEL experiment**

The experimental setup is shown in Fig. 2a. The SwissFEL and the Bernina beamline layout are described elsewhere [*Ingold2019*]. Since the experiment was performed as part of the commissioning for the facility, the X-ray photon energy was limited to 2.2 keV. The third harmonic (6.6 keV) was used to probe the structural changes after photoexcitation. The energy per pulse was ~0.2mJ and the third harmonic content of about 1%. The fundamental was suppressed using absorbers. We estimate about $10^9$ 3rd harmonic photons per pulse at the sample position. The X-rays were focused by a Kirkpatrick-Baez mirror system. The spot was set to ~ 200 x 7.5 µm² (FWHM h x v). Because of the small grazing incidence angle of 0.5° the small vertical spot was essential to minimize the footprint on sample which with a length of 0.86 mm footprint is still one of the major contributions to the peak broadening. The pump laser pulses, generated by a Coherent CPA amplifier (Legend) were purposely stretched to 0.5 ps to limit the peak intensity on the sample increasing the overall time resolution to ~600 fs. Note that no jitter correction [*Harmand2013*] was used for the experiment. The measured instrumental time resolution of 350 fs FWHM [*Ingold2019*] increased due to the pump pulse stretching to about 600 fs. The laser beamsize at the sample position was 300 um (FWHM). With an incident angle of 10° the resulting footprint was 300 x 730 µm². The repetition rate of the laser was set to 5 Hz minimizing residual heating effect.

**Data reduction/correction**

Diffracted X-rays were measured (for every single pulse) by the Jungfrau pixel detector [*Mozzanica2018*]. The detector was calibrated to convert ADU directly in equivalent keV photon energy. Most pixels (> 70%) had zero counts, the others detected either one elastic photon (~6.6 keV reading, about 3% of the pixels), a Ti fluorescent photon (centered around 4.4 keV, 11% of the pixels) or a combination of the two (1 elastic +1 fluorescence photon). By eliminating the fluorescence photons from each image a lower background image was obtained. Each "fluorescence corrected" image has been azimuthally integrated to obtain 1D curve [*Ashiotis2015*]. Each data point was collected by acquiring 500 images (250 with pump laser and 250 without pump laser). The two sets were averaged (after "fluorescence correction" and azimuthal integration) to provide two curves (laser off and on). To correct for drifts in photon energy each "off image" was used to extract the average photon energy during collection of a given time delay. The corresponding "laser on" image was treated using the optimized X-ray photon energy. To further verify the stability of the data collection and correction strategy, every 10 delays a reference time delay was acquired to be sure that no drifts or permanent sample change had happened. Each powder pattern was normalized to the intensity of the air scattering contribution dominating at low scattering vectors (between 0.7 and 0.9 Å$^{-1}$). All raw data and data reduction scripts will be made available.

**Data analysis (Rietveld refinement)**

Rietveld whole powder pattern profile refinements were performed following an essentially fundamental parameter approach using the TOPAS software [*Coelho2018*]. Pawley refinements were also used and both methods gave consistent evolutions of the unit cell parameters and scales factors. The profiles were described using a beam energy of 6.5 keV ($\lambda$ = 1.899 Å) and a gaussian emission profile with 1.3% FWHM. The peak width and sample displacement were described using the expressions from [*Rowles2017*], vertical width and incident angle of the beam being fixed as measured (4 µm and 0.5° respectively). The sample detector-distance was fixed to 61 mm, as refined in the azimuthal integration step. The absorption coefficient at 6.5 keV was calculated as 800 cm$^{-1}$ (for a packing density of 0.8, measured from x-ray transmission measurement). These parameters were fixed for the refinements against TR-patterns. The refinements included two phases ($\beta$- and $\lambda$-). For each phase, the free parameters were the cell parameters (a, b, c, $\phi$), the atomic position in the (***a***, ***c***) plane for the five independent atoms

(position along **b** being symmetry-restricted) and the scale factor. An additional microstrain-type gaussian convolution (FWHM = $\varepsilon$ x tan($\theta$)) accounted for the strain distribution observed during the propagation (Fig. S7). The initial Bragg peak profile was well defined by the experimental resolution function defined as described above. Thus no extra size contribution was considered in the refinement of the reference patterns. The transformation being reversible and isostructural, no evolution of the crystallite size was expected and thus considered either. In any case, the choice was made to retain the simplest convolution functions to ensure the robustness of the refinement. Hence no lorentzian contribution, asymmetry or anisotropy was taken into account. The background was described as a 3$^{rd}$ order polynomial. The texture of the majority phase was described using Spherical Harmonics corrections, whose coefficients were refined on reference patterns and then kept constant for all delays. The $R_{wp}$ agreement factors were around 7.5%; $R_{Bragg}$ were calculated around 2% for the majority $\beta$ - phase and around 7% for the $\lambda$ - phase. For the SwissFEL data refinement, errors were estimated from the distribution (standard deviation *SD*) of the refined values obtained on reference pattern (interleaves data with no laser), and thus have to be understood as relative errors; the errors given in the text and in the figures corresponds to *2xSD*. Note that for the ESRF data, the quality of the extracted patterns did not allow to refine the atomic positions. In both cases, reference structures were taken from [**Ohkoshi2010**, **Onoda1998**] and atomic displacement parameters were fixed to the equilibrium room temperature value due to limited accessible q region.

# Figures

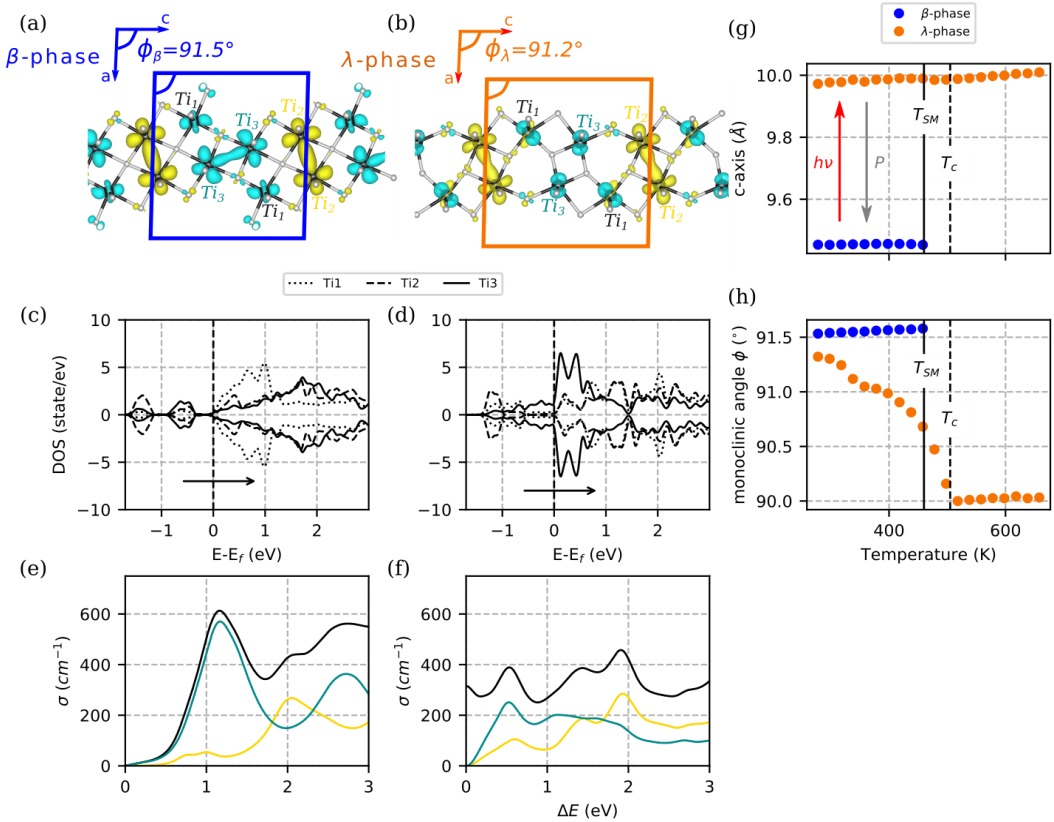

*Figure 1 : Structural and optical changes upon phase transition in $Ti_3O_5$ nanocrystals at thermal equilibrium* **a**) and **b**) atomic structure of $\beta$ - and $\lambda$ - phases, respectively, with calculated charge density for the two bands centered at 1.1 (yellow) and 0.3 eV (blue) below the Fermi level, strongly localized around $Ti_2$-$Ti_2$ and $Ti_3$-$Ti_3$ dimmers; **c**) and **d**) Calculated density of state projected on $Ti_1$(dotted lines), $Ti_2$(dashed lines) and $Ti_3$ (plain lines) for $\beta$ - and $\lambda$ - phases, respectively. The black arrow stands for the energy transfer corresponding to the 1.55 eV pump photons.; **e**) and **f**) Mean calculated Optical conductivity for $\beta$ - and $\lambda$ - phases, respectively (black line). Yellow (blue) curve represents the contribution arising from the band at 1.1 eV (0.3 eV). Diagonal contributions are shown in Fig. S1; **g**) and **h**) X-ray powder diffraction study on $Ti_3O_5$ nanocrystals: **g**) change in c-parameter of the monoclinic unit cell revealing with a jump which underlines the strongly first order phase transition at $T_{SM}$ = 460 K between $\beta$ - and $\lambda$ -phases ; **h**) Evolution with temperature of the monoclinic angle $\phi$ exhibiting a jump upon the first order phase transition at $T_{SM}$ = 460 K (between $\beta$ - and $\lambda$ -phases which co-exist in these nanocrystals below $T_{SM}$), and a continuous locking at 90 ° characterizing the symmetry change towards *Cmcm* high temperature $\alpha$ -phase. The critical temperature $T_c$ = 500 K for this second order phase transition is pointed out with a plain vertical line.

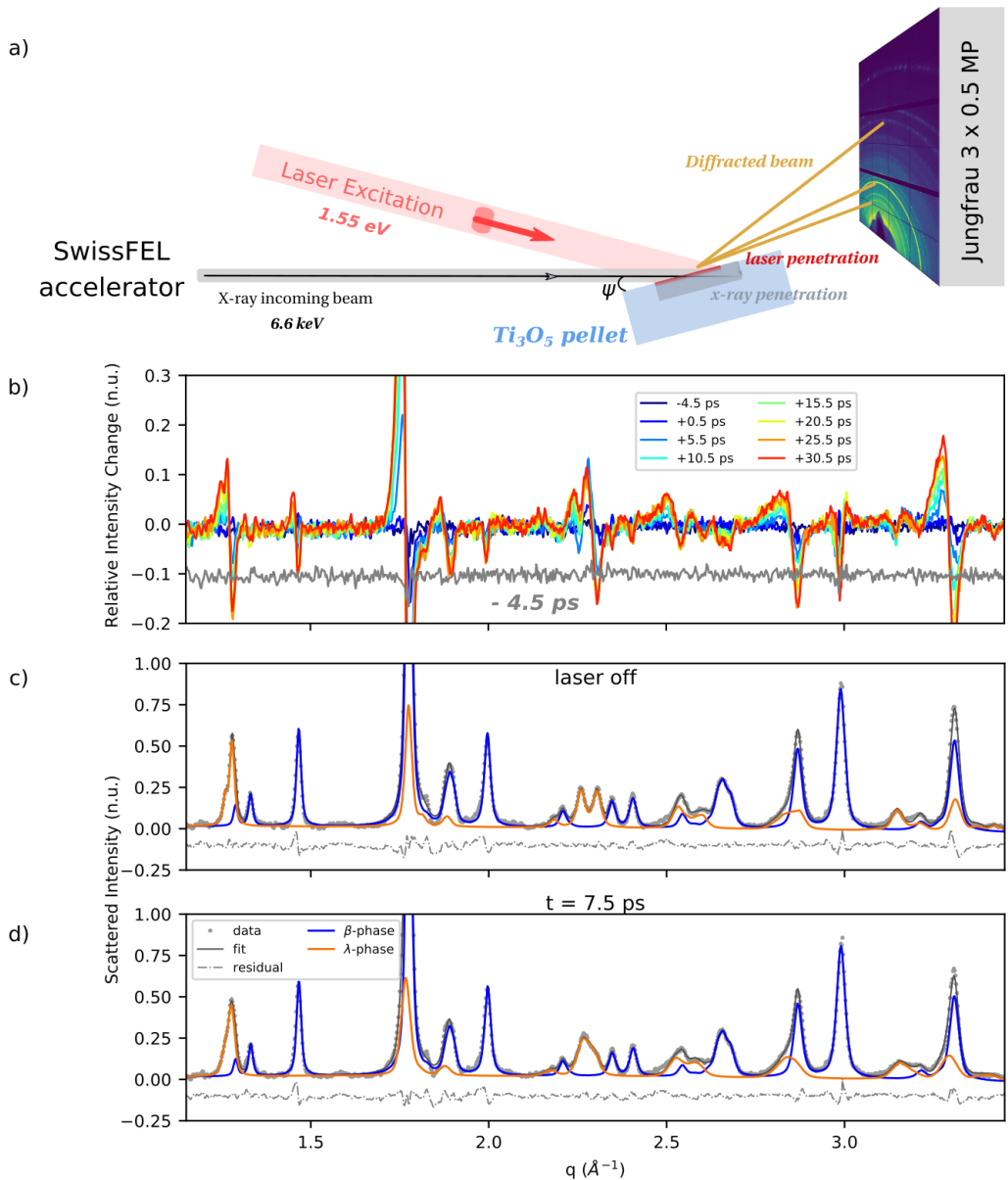

*Figure 2: Experimental setup, raw data and structural refinement.* **a**) Experimental setup for time resolved powder X-ray diffraction in quasi grazing angle geometry. The Debye-Scherrer rings are collected on a 2D Jungfrau fast readout detector with single photon sensitivity [***Mozzanica2018***]. **b**) Differential patterns *(laser$_{on}$-laser$_{off}$)* showing up to 30% variations of the signal (pump fluence 0,85 mJ.mm$^2$) on ps time scales ( < 35 ps ). Grey curve shows negative -4.5 ps delay for reference (shifted by -0.1 along y for clarity). **c**) and **d**) Rietveld refinement of reference spectra (with no laser) and spectra at t = 7.5 ps respectively. Measured powder patterns are plotted in light grey, plain circles and result of Rietveld refinement in black, plain line. Orange and blue patterns are respective contributions to the refinement of the $\lambda$ - and $\beta$ - phases; the residual curve is shown in grey, dashed lines.

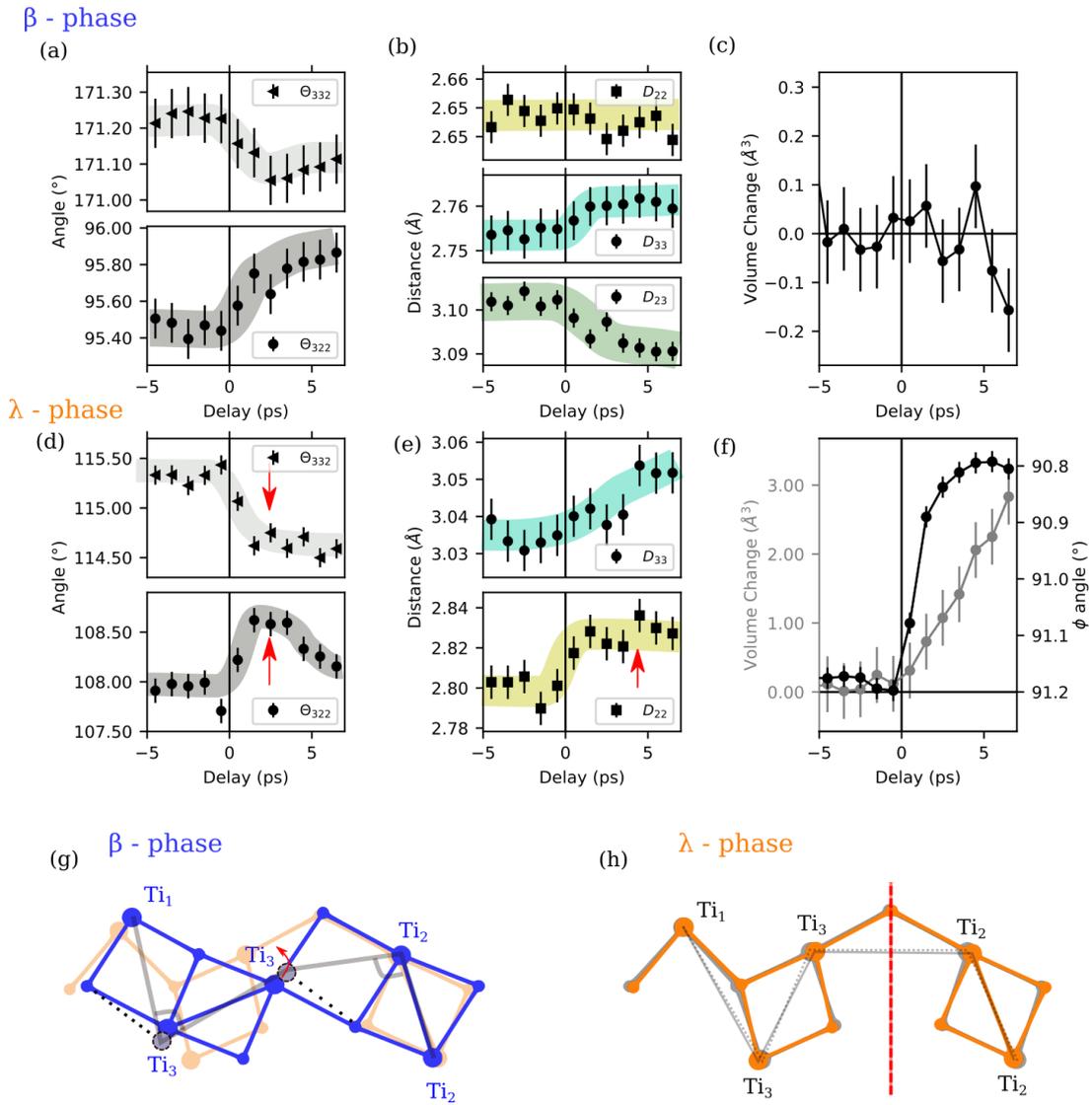

*Figure 3: Ultrafast structural changes extracted from X-ray diffraction data.* **a-f**: Evolution of relevant inter-atomic angles, distances and unit cell volume change for the $\beta$ - (**a-c**) and $\lambda$ - (**d-f**) phases ($\theta_{ijk}$ = $Ti_i$-$Ti_j$-$Ti_k$ angles and $D_{ij}$ = $Ti_i$-$Ti_j$ distances). The evolution of the monoclinic $\phi$ angle is also reported in the case of the $\lambda$ - phase (**f**). Thick transparent solid colored lines are guides for the eye. **g-h**: schematic representation of the observed local distortions (exaggerated for clarity) for the $\beta$ - (**g**) and $\lambda$ - (**h**). in (**g**), the $\lambda$ - structure is represented in light orange for reference and in (**h**), the red line is the *m* mirror planes lost at $T_c$.

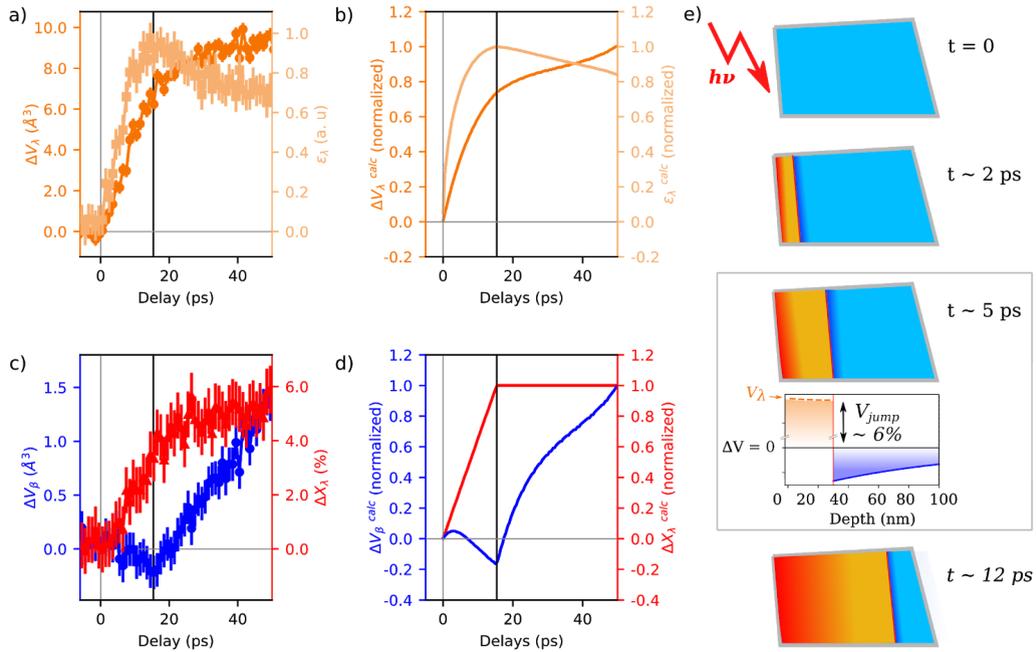

*Figure 4 : Evolution of long range order and phase change on the acoustic time scales:* **a)** and **c)** Parameters extracted from Rietveld refinement of TR-XRD data. **a)** Temporal evolution of $\lambda$-phase unit cell volume ($\Delta V_\lambda$, dark orange, plain circles) and microstrain parameter $\varepsilon_\lambda$ (light orange, plain squares). **c)** Temporal evolution of $\beta$-phase unit cell volume ($\Delta V_\beta$, blue) and $\lambda$-phase fraction ($\Delta X_\lambda$, red). Relative changes normalized to the change at 50 ps and errors estimated as described in the Methods section. **b)** and **d)** Simulation of respectively **a)** and **c)** with Thomsen model [*Thomsen1986a*]. All details are given in the SI. **b)** $\lambda$-crystallites. Calculation of expected evolution of the volume change ($\Delta V_\lambda^{calc}$, dark orange) and microstrain ($\varepsilon_\lambda^{calc}$, light orange) following ultrafast excitation. Parameters of the model are taken as follows: laser penetration depth of $\xi_L$ = 65 nm, sound velocity of $V_s$ = 6.5e3 m.s$^{-1}$; curves normalized to the value at t = 50 ps. **d)** $\beta$-crystallites. Calculation of expected evolution of the volume change ($\Delta V_\beta^{calc}$, blue) and $\lambda$-phase fraction ($\Delta X_\lambda^{calc}$, red) following ultrafast excitation. The corresponding acoustic front propagation time corresponding to the mean crystallite length (L = 100 nm, $\tau_{PF}$ = L /$v_s$) is shown with solid black lines in **a)**, **b)**,**c)** and **d)**. **e)** Schematic representation of phase front propagation within one crystallite and associated volume deformation calculated at 5 ps (arbitrary scale); blue for $\beta$-phase, orange for $\lambda$-phase; colour gradients illustrates strain gradient in each phase.

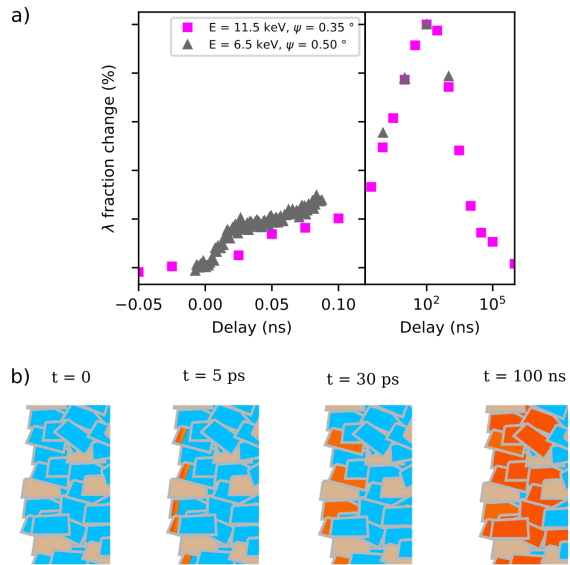

*Figure 5: Thermal propagation in the bulk pellet.* **a**) Multiscale evolution (from ps to ms) of the $\lambda$ - phase fraction as extracted from TR-XRD measurements at the ESRF beamline ID09 (magenta squares). The fraction change is normalized to the maximum value for allowing a direct comparison with the evolution observed at SwissFEL Bernina (grey triangles). Long time delays are presented on a logarithmic time scale. **b**) Schematic view of the proposed phase transition propagation mechanism. The initial pellet is composed of $\lambda$ - (in grey) and $\beta$ - (in blue) crystallites. Photo-induced precursor distortions launch an acoustic wave starting from the pellet surface transforming the $\beta$ - crystallites lying on the surface (0 - 20 ps, see Fig. 4). Heat propagation causes late structural changes that extend up to 200 nm from the surface when the transformation reaches its maximum (100 ns).

# Supplementary Figures

## Optical conductivity calculations

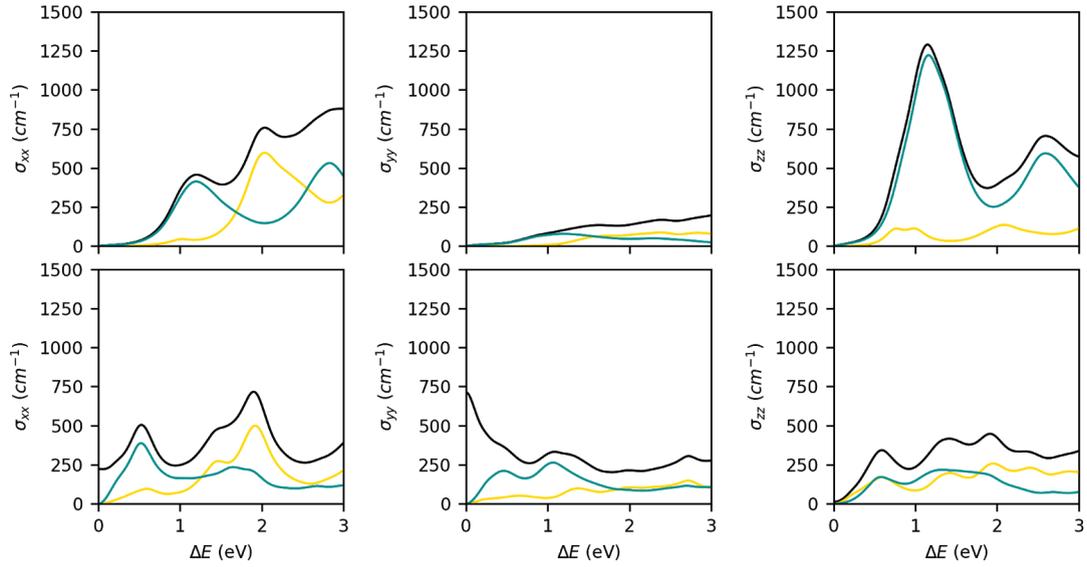

**FigureS1**: Diagonal contributions of the optical conductivity tensor for the β- (top) and $\lambda$-(bottom) phases of $Ti_3O_5$, calculated along the x, y, z direction of the primitive unit cell. Total calculated conductivity in black; Yellow (blue) curve represents the contribution arising from the band at 1.1 eV (0.3 eV) below Fermi level.

## Surface Morphology

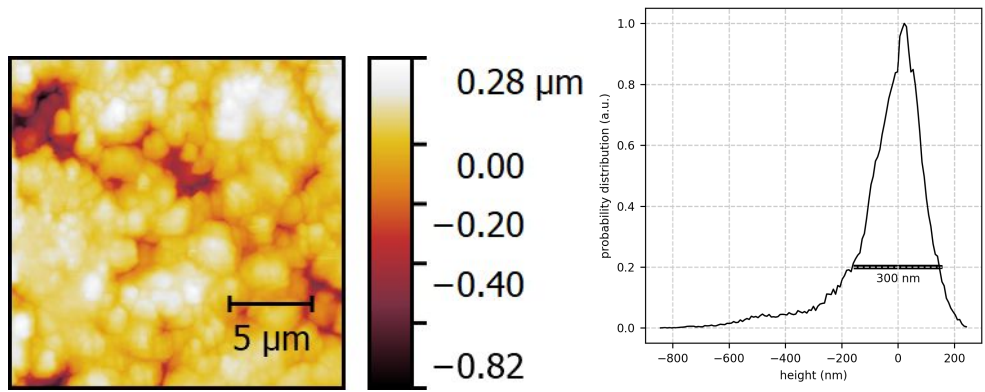

**Figure S2** : 20 x 20 um² AFM height image of $Ti_3O_5$ pellet surface revealing the surface roughness with typical peak to valley amplitude around 1 µm. Such roughness limits the surface sensitivity of the X-ray experiments as discussed in the Methods. Measurement on the same pellet as that used for the SwissFEL experiment.

## Angle dependent penetration depth for a perfect surface

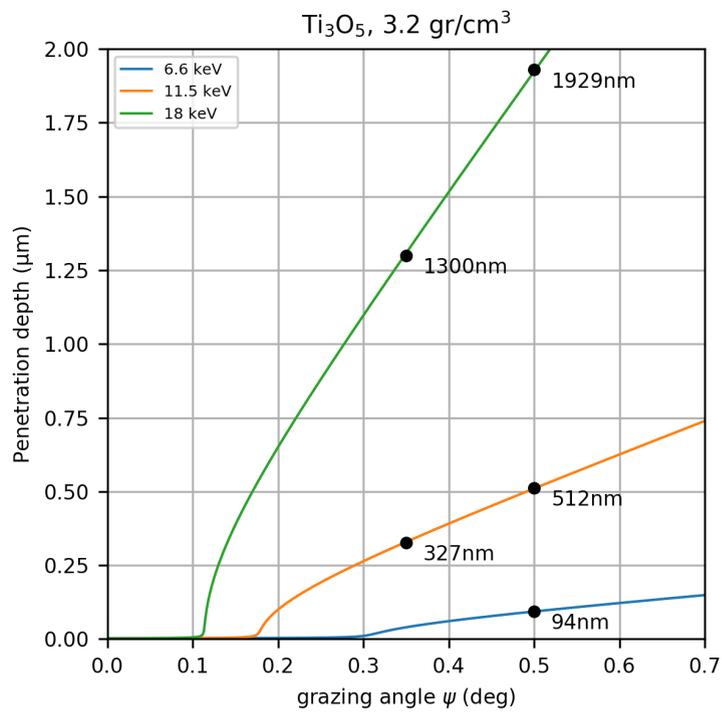

**Figure S3**: Dependence of the X-ray penetration depth as a function of photon energy and grazing angle, calculated using the [**Henke1993**], for a $Ti_3O_5$ sample of density measured for pellet. Black circles for conditions shown in Fig. S4.

## Long time scale dynamics

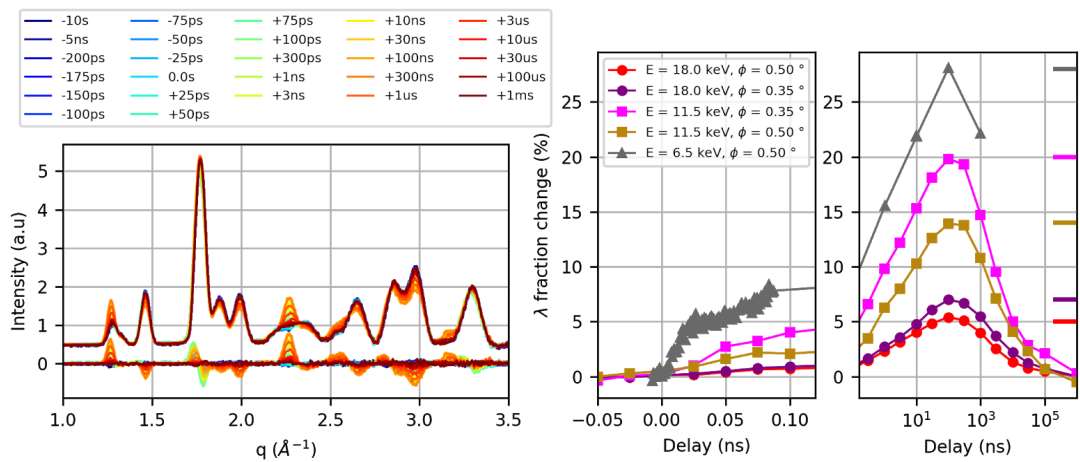

**Figure S4**: Evolution of the $\lambda$ - phase fraction from ps to ms time-scales as extracted from TR-XRD data measured on beamline ID09 at ESRF. Right panel) Effect of x-ray probe energy and incidence angle at fixed pump power density (1.84 mJ/mm$^2$). Data emphasise decreased X-ray penetration into pellet at smaller incidence angle or x-ray energy, hence at fixed laser power average $\lambda$ - phase fraction increases as described in S9. The Swissfel data are also shown for comparison (grey triangles). The level bars on the right axis represent the expected fraction of photoinduced $\lambda$ phase, calculated according to the procedure described in the SI paragraph "Probed Penetration Depth for Ti$_3$O$_5$ pellet and switching efficiency"; colours are matched with data. Left panel) Typical time-resolved absolute and differential patterns measured at ID09 (x-ray energy = 11.5 keV, power density = 1.84 mJ/mm$^2$).

# Rietveld refinement

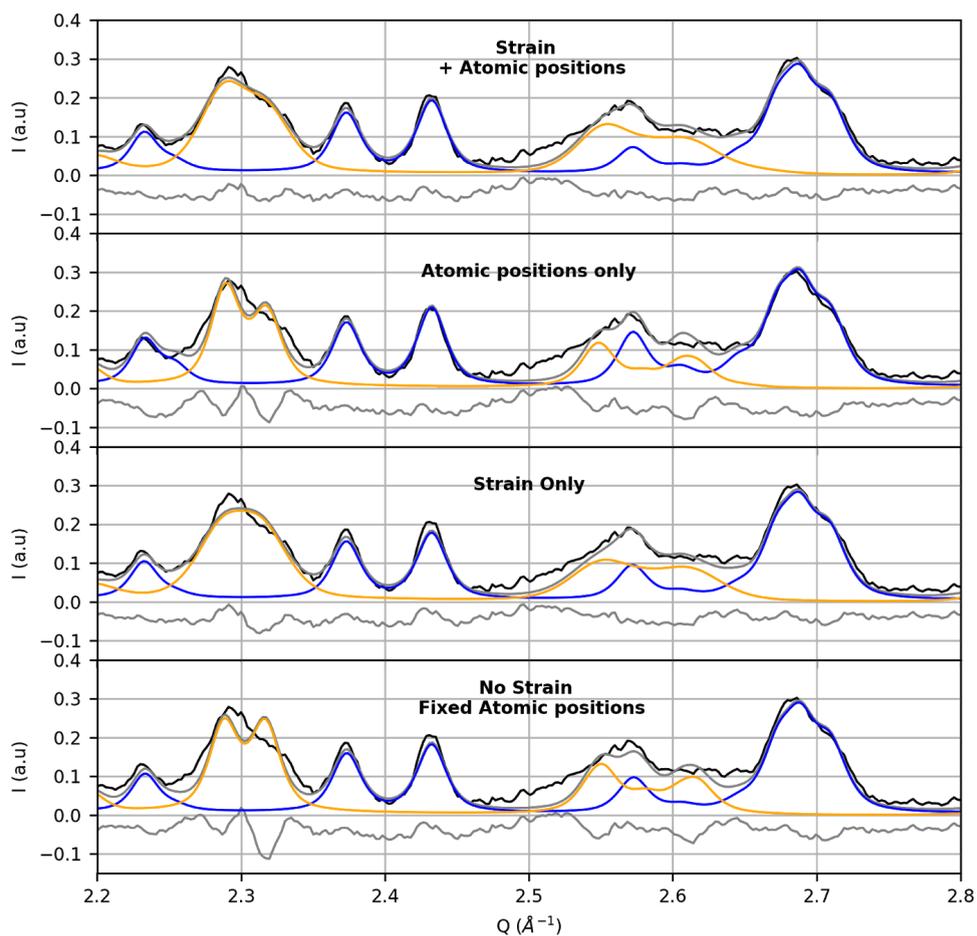

**Figure S5**: Quality comparison between different sets of refined parameters, for a diffraction pattern at t = 7.5 ps. In all sets, unit cell parameters a, b, c, $\phi$ are refined. Experimental patterns are shown in black, refined patterns in grey, contributions of $\beta$- and $\lambda$ - phase in blue and orange respectively, residual values in grey. From bottom to top: atomic positions fixed and microstrain not considered, microstrain refined, only atomic positions refined, atomic positions and microstrain refined.

## Diffuse scattering measured of single crystals

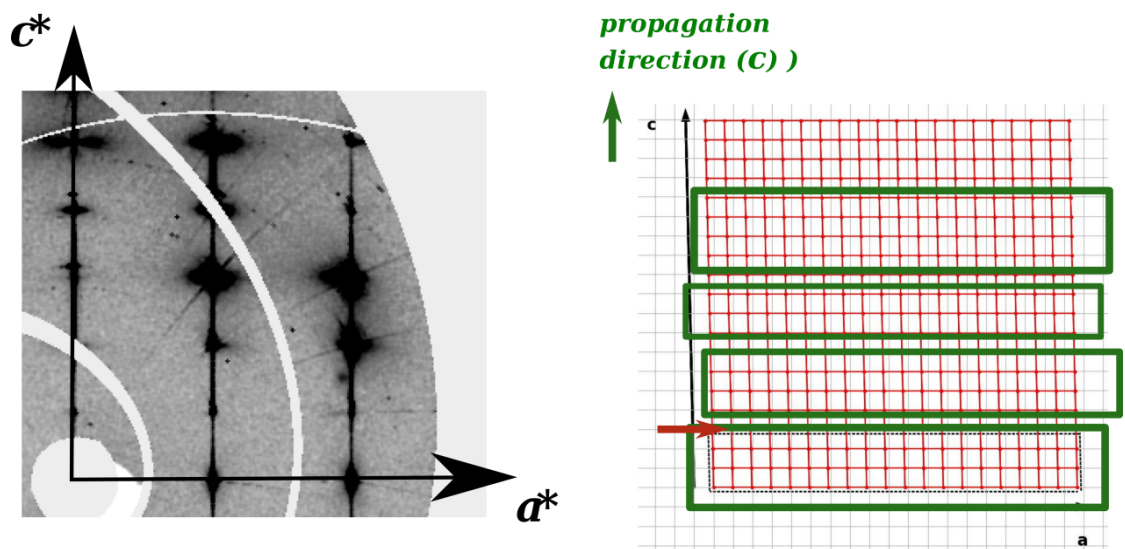

**Figure S6**: left : ($a^*$, $c^*$) reciprocal plane reconstructed from X-ray diffraction measurements at room temperature on $Ti_3O_5$ single crystals (360° rotation, 0.1° per images, shutterless mode). Measurement was performed on ID28 beamline at ESRF on a 4-circle diffractometer equipped with pilatus 2M detector. X-ray photon energy was 26 keV. Contrast is enhanced to highlight strong diffuse scattering lines lying along the *c\** axis, at integer position along *a\**. These lines indicate stacking faults propagating along *c*, which might arise from the ferroelastic distortion as shown schematically on the right. The exact coherence length could not be quantified precisely, but the intensity suggests only a few unit cells (green rectangle on the right scheme), compared with analogous observations for inorganic crystals [**Zhong2001**].

# Model calculations of Strain, Volume change and Microstrain

Phenomenological equations were derived from the Thomsen model [**Thomsen1986a**].

This model considers an initial thermal stress profile of the form:

$$R_e \times R_T \times A e^{-z/\xi} \quad (eq.1)$$

Where $R_e = -3B\beta$ is the thermoelastic response of the material, B stands for the bulk modulus and $\beta$ for the linear expansion coefficient.

$R_T = 1/C$ with C the heat capacity

$A = (1-R)\frac{Q}{F\xi}$ is the effective laser power density absorbed by the sample (R : optical reflectivity, Q : laser power, F : laser footprint on the sample surface; $\xi$ : laser penetration depth)

The strain is then given by the following expression:

$$\eta(z, t) = R_e \times R_T \times A \times P \times \left[ e^{-z/\xi} \left(1 - \tfrac{1}{2} e^{-v_s t/\xi}\right) - \tfrac{1}{2} e^{-|z - v_s t|/\xi} \times sign(z - v_s t) \right] \quad (eq.2)$$

Where $P = \frac{1+v}{1-v}$ with $v$ the poisson coefficient.

In our case, the initial stress due to the laser pulse is assumed to include both initial T increase and local precursor distortions induced by the photo-excitation. For each phase we assumed a thermoelastic contribution to strain in the form:

$$\eta_T(z, t) = S \times \left[ e^{-z/\xi} \left(1 - \tfrac{1}{2} e^{-v_s t/\xi}\right) - \tfrac{1}{2} e^{-|z - v_s t|/\xi} \times sign(z - v_s t) \right] \quad (eq.3)$$

Where $S = R_e \times R_T \times A \times P$

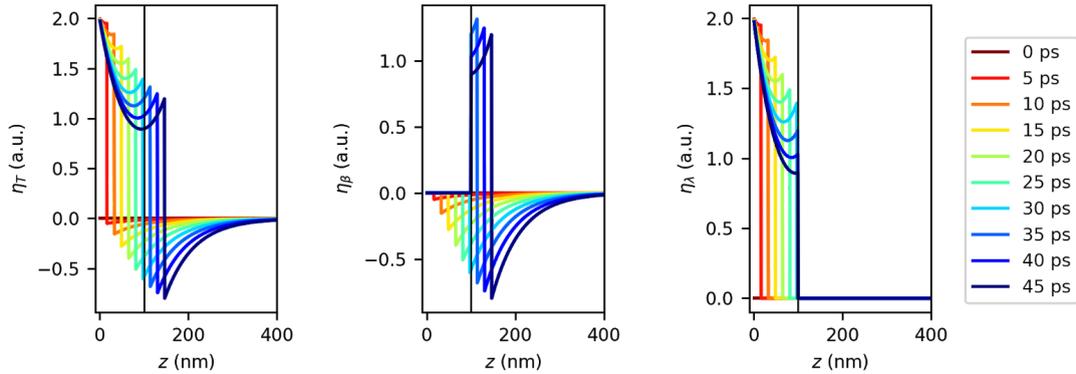

**FigureS7**: Strain contributions calculated for different time delays (from 3 ps to 50 ps, colour graded from red to dark blue, and from left to right: total $\eta_T(z, t)$, $\beta$ phase $\eta_\beta(z, t)$ and $\lambda$-phase $\eta_\lambda(z, t)$. The mean crystallite length (L = 100 nm) is shown with solid black lines.

The phase transition was taken into account as follows:

- For the transformed β crystallites, the strain contribution was written as follows:

$$\eta_\beta(z, t) = s_\beta \times S_\beta \times \left[ e^{-z/\xi} \left(1 - \tfrac{1}{2}e^{-v_s t/\xi}\right) - \tfrac{1}{2}e^{-|z-v_s t|/\xi} \times sign(z - v_s t) \right] \quad (eq.4)$$

with:

$s_\beta = 0$ for $z - v_s t < 0$ and $z < z_S$ (accounts for β > λ transition)

$s_\beta = 1$ for $z - v_s t > 0$ or $z > z_S$

With $S_\beta$ taken arbitrarily higher than $S$ : $S_\beta/S = 3$ (best match with experimental data, accounting for the fact that the λ phase has a much higher volume that the β phase) and $z_S = 100$ nm the depth where the phase front is assumed to stop.

- For the newly formed λ crystallites, the strain contribution was written as follows:

$$\eta_\lambda(z, t) = s_\lambda \times S_\lambda \times \left[ e^{-z/\xi} \left(1 - \tfrac{1}{2}e^{-v_s t/\xi}\right) - \tfrac{1}{2}e^{-|z-v_s t|/\xi} \times sign(z - v_s t) \right] \quad (eq.5)$$

with:

$s_\lambda = 1$ for $z - v_s t < 0$ and $z < z_S$

$s_\lambda = 0$ for $z - v_s t > 0$ or $z > z_S$ (accounts for λ < β transition)

With $S_\lambda$ taken arbitrarily smaller than $S$ : $S_\lambda/S = 0.05$ (best match with experimental data, accounting for the fact that the newly formed λ phase is less dilated as the initial excited one)

- For each phase, we assumed that the contributions are independent, so that the total strain is simply calculated as the weighted sum of the two contributions.

$$\eta_{\beta_{tot}}(z, t) = r \times \eta_\beta(z, t) + (1 - r) \times \eta_T(z, t)$$

$$\eta_{\lambda_{tot}}(z, t) = r \times \eta_\lambda(z, t) + (1 - \eta) \times \eta_T(z, t)$$

Were r is the estimated ratio of transformed β - crystallites over the first $z_S = 100$ nm at the end of the phase front propagation (0.26, see next paragraph).

The strain and standard deviation are then integrated between $z = 0$ (surface) and $z_p = 400$ nm, (estimated penetration depth of the Xrays) (see Fig. 4 b and d in the main text).

$$Z_{ph} = \int_{z=0}^{z_P} \eta_{ph_{tot}}(z, t) dz$$

$$S.D._{ph} = \sqrt{1/z_P \times \int_{z=0}^{z_P} \left(\eta_{ph_{tot}}(z, t) - \overline{\eta_{ph_{tot}}(z, t)}\right) dz}$$

With $ph = \beta, \lambda$.

The calculated uniaxial distortion leads to a global volume change of the excited sample by the following relation:

$$\Delta V^{calc} = A * Z$$

$$\varepsilon^{calc} = \sqrt{A} * S.D.$$

$\Delta V$ and $\varepsilon$ being proportional to the change of the unit cell volume and microstrain (respectively) extracted from Rietveld refinement against experimental patterns.

Assuming that the observed ratio is average over the x-ray penetration depth $z_P$ = 400 nm, the $\lambda$ - fraction $X_\lambda(t)$ at time t is given by:

$$X_\lambda^{calc}(t) \times z_P = X^* min(z(t), z_S) + X_0(z_P - min(z(t), z_S))$$

$$X_\beta^{calc}(t) = 1 - X_\lambda^{calc}(t)$$

where $X^*$ is the new fraction after phase front and $X_0$ is the initial fraction before excitation, supposed homogeneous, $z_S$ is the depth where the phase front propagation stops (100 nm)

## Picosecond Interferometry measurement

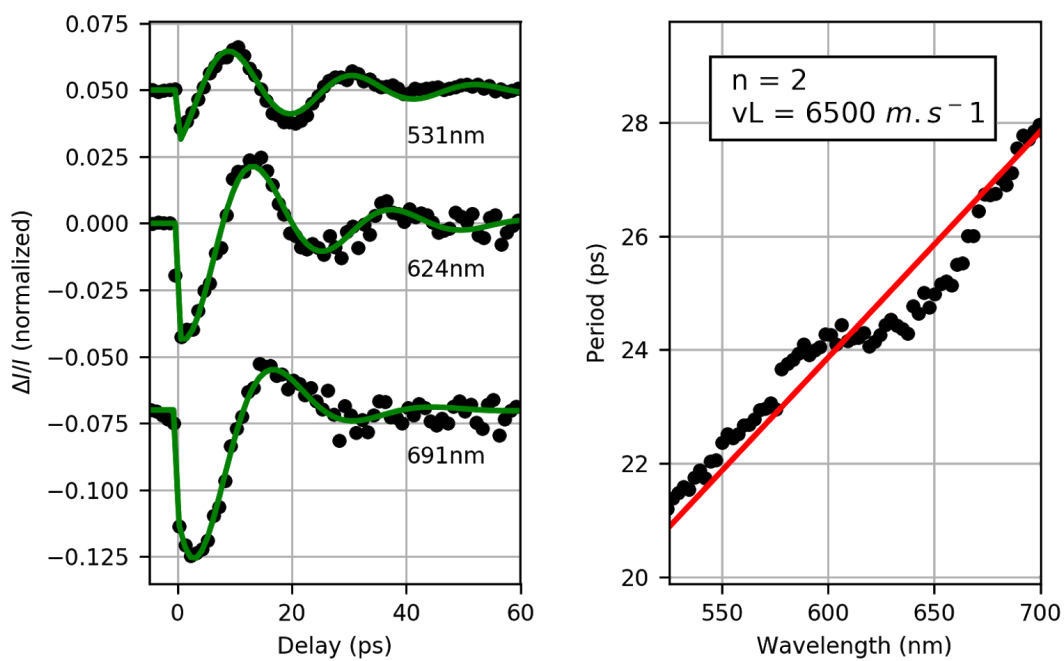

**Figure S8**: Time resolved reflectivity measured on Ti$_3$O$_5$ single crystal with visible pump / white light probe setup at IPR , pump set to 1.55 eV. Left: oscillatory part of the time dependent reflectivity for selected wavelengths of white light probe illustrating oscillation period dependence on wavelength. Right: oscillation period extracted from data refinement. The linear dependence of period was refined following the model proposed in [**Thomsen1986b**]. Extracted periods are in agreement with those previously reported for a pellet sample [**Asahara2014**], and with real part of optical index from [**Hakoe2017**], they yield sound velocity of 6.5e3 m.s$^{-1}$.

# Probed Penetration Depth for $Ti_3O_5$ pellet and switching efficiency

This section aims at obtaining the switching efficiency from the experimentally determined one. The latter one is influenced by the X-ray penetration depth. For ideal surfaces, the X-ray penetration depth can be calculated using tabulated values [**Henke1993**] as implemented using the CXRO website http://henke.lbl.gov/optical_constants/atten2.html. The results for few X-ray photon energies and angles are given in Fig. S3. For sufficiently low angles total external reflection results in an extremely small penetration depth (few nm). The calculated values for the X-ray photon energies and grazing angles shown in Fig. S4 are given in Fig. S3. They range from 92 nm (SwissFEL experiment, 6.6 keV, $\psi$ = 0.5 deg) to 1.93 um (ESRF experiment, 18 keV, $\psi$ = 0.5 deg). The pellet granularity also determines the effective probed depth. The pellets have a typically roughness of 300nm (Fig. S2). The combined effect of the roughness and ideal penetration depth results in what we call "effective penetration depth" ($z_p$ in the equations below). Since the layer converted by the strain wave (~100 nm) is smaller than the effective penetration depth (> 300 nm) a smaller apparent phototransformed fraction will be observed.

By using a simplified model shown in Fig. S9 we can calculate the apparent fractions as a function of relevant physical parameters. In particular we assume that strain wave propagation and heat diffusion processes transform a finite fraction of the $\beta$ phase, and the respective efficiencies are denoted $\varepsilon_s$ and $\varepsilon_h$.

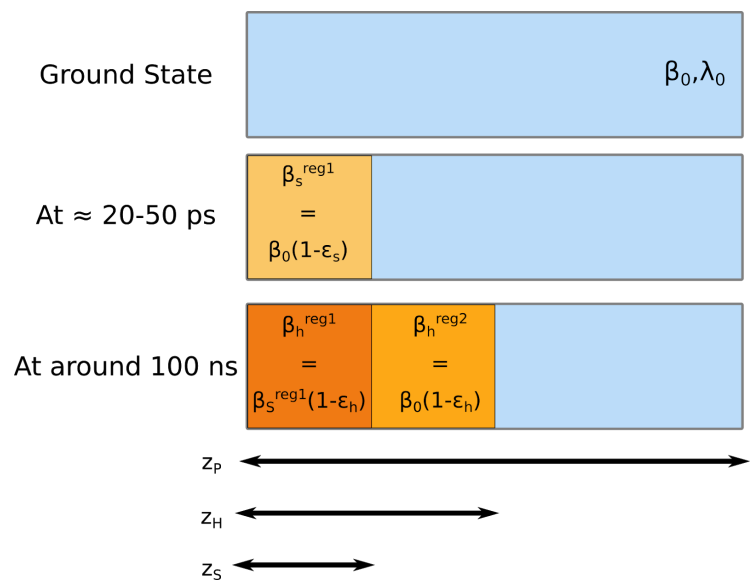

Figure S9: Model used to calculate the switching efficiencies. The horizontal axis represents the sample thickness. Without photoexcitation the sample is in a mixture of lambda and beta phases indicated respectively as $\lambda_0$ (~0.25) and $\beta_0$ (~0.75). After the strain wave propagation a fraction of the beta phase will be converted to lambda with a certain efficiency ($\varepsilon_s$). At longer time extra switching is observed due to thermal transition. In each region the increase of lambda phase is assumed to be equal to the remaining beta fraction in that region times the efficiency of "heating switching" ($\varepsilon_h$). Summing the beta and lambda phase fraction at each step and in each region allows to calculate the expected measured valued.

These models results in simple expressions that link the physical parameters to the observed change of fraction after ~20 ps and at ~100 ns:

$$\Delta f_s^\lambda = \beta_0 \varepsilon_s \frac{z_S}{z_P}$$

$$\Delta f_h^\lambda = \beta_0(1-\varepsilon_s)(1-\varepsilon_h)\frac{z_S}{z_P} + \beta_0(1-\varepsilon_h)\frac{z_H-z_S}{z_P} + \beta_0\frac{z_P-z_H}{z_P}$$

With $z_s, z_h, z_p$ are the strain wave propagation (100 nm), heat diffusion (200 nm) and probed depth respectively. Note that the fraction of $\lambda$ phase (referred to as $X$ in the previous section) can also be written as $X = \lambda_0 + \Delta f_s^\lambda$.

Assuming $\varepsilon_s$ and $\varepsilon_h$ to be 0.26 and 0.7, respectively, this results in values reported in the table below. The calculated values are obtained from the above expression, the experimental ones from Fig. S4.

|  | E (keV) | grazing angle $\psi$ (°) | $z_P$ (nm) | $\Delta f_s^\lambda$ (calc) | $\Delta f_s^\lambda$ (exp) | $\Delta f_h^\lambda$ (calc) | $\Delta f_h^\lambda$ (exp) |
|---|---|---|---|---|---|---|---|
| SwissFEL | 6.6 | 0.5 | 394 | 0.048 | 0.05 | 0.27 | 0.28 |
| ESRF | 11.5 | 0.35 | 627 | 0.030 | 0.03 | 0.17 | 0.20 |
|  | 11.5 | 0.50 | 812 | 0.023 | 0.025 | 0.13 | 0.14 |
|  | 18.0 | 0.35 | 1600 | 0.011 | 0.01 | 0.07 | 0.07 |
|  | 18.0 | 0.50 | 2229 | 0.008 | 0.01 | 0.05 | 0.05 |